\documentclass[useAMS,usenatbib]{mn2e}
\usepackage{graphics,graphicx,array,amssymb,times,fontenc,mathptm,amsmath}

\usepackage{supertabular}

\title[Variable stars in the Magellanic Clouds]{Variable stars in the Magellanic Clouds: \\ II. The data and infrared properties}
\author[Y. Ita et al.]{Yoshifusa Ita$^{1,2}$\thanks{E-mail: yita@ir.isas.jaxa.jp}\thanks{On leave from the University of Tokyo}, Toshihiko Tanab\'{e}$^{1}$, Noriyuki Matsunaga$^{1}$, Yasushi Nakajima$^{3}$, Chie
\newauthor Nagashima$^{3}$, Takahiro Nagayama$^{4}$, Daisuke Kato$^{4}$, Mikio Kurita$^{4}$, Tetsuya Nagata$^{4}$,
\newauthor Shuji Sato$^{4}$, Motohide Tamura$^{3}$, Hidehiko Nakaya$^{5}$ and Yoshikazu Nakada$^{1,6}$
\\
$^1$Institute of Astronomy, School of Science, The University of Tokyo, Mitaka, Tokyo 181-0015, Japan\\
$^2$Institute of Space and Astronautical Science, Japan Aerospace Exploration Agency, Sagamihara, Kanagawa 229-8510, Japan\\
$^3$National Astronomical Observatory of Japan, Mitaka, Tokyo 181-8588, Japan\\
$^4$Department of Astrophysics, Nagoya University, Chikusa-ku, Nagoya 464-8602, Japan\\
$^5$Subaru Telescope, National Astronomical Observatory of Japan, 650 North A'ohoku Place, Hilo, HI 96720, U.S.A.\\
$^6$Kiso Observatory, School of Science, The University of Tokyo, Mitake, Kiso, Nagano 397-0101, Japan
}
\begin{document}

\date{Received -- / Accepted --}

\pagerange{\pageref{firstpage}--\pageref{lastpage}} \pubyear{2003}

\maketitle

\label{firstpage}

\begin{abstract}The data of 8,852 and 2,927 variable stars detected by the OGLE survey in the Large and Small Magellanic Clouds are presented. They are cross-identified with the SIRIUS $JHK$ survey data, and their infrared properties are discussed. Variable red giants are well separated on the period-$J - K$ plane, suggesting that it could be a good tool to distinguish their pulsation mode and type.
\end{abstract}


\section{Introduction}
Because of their proximity and comparatively well-known distances, the Large and Small Magellanic Clouds (LMC and SMC, respectively) are the ideal places to study stellar evolution. To search for the gravitational lensing events, several projects are now underway monitoring the sky toward the Magellanic Clouds and also the Galactic bulge, where there are a large number of background stars that are potential targets for microlensing. As the natural by-products of these surveys, many variable stars were newly found. The Optical Gravitational Lensing Experiment (OGLE) project is one of such surveys and its $I$ band time-series data in the Magellanic Clouds (\citealt{zebrun2001}) obtained during four years of OGLE-II run (\citealt*{udalski1997}; January 1997 - November 2000) is now available over the Internet (OGLE homepage; http://sirius.astrouw.edu.pl/$^{\sim}$ogle/).

To detect dust-enshrouded variables that have possibly escaped from the
ongoing/previous optical projects, we started the $JHK$ monitoring
survey toward the Large and Small Magellanic Clouds using the InfraRed
Survey Facility (IRSF) at the South African Astronomical Observatory
Sutherland station. The IRSF consists of a dedicated 1.4m telescope to
which is attached a near infrared (NIR) camera SIRIUS, which can observe
the sky in the three wave bands ($JHK_s$) simultaneously. The filter characteristics are
shown in table~\ref{filter}. The field of view of this system is about
7.7$^\prime$$\times$7.7$^\prime$ with a scale of
0.453$^{\prime\prime}$/pixel. Details of the instrument are found in \citet{nagashima} and \citet{nagayama}. This campaign was initiated in December 2000, and so far, many dust-enshrouded variables were found and their light variation data has been steadily accumulating (see \citealt{ita} for the first results).

\begin{table}
\caption{The filter characteristics.}
\label{filter}
\centering
\begin{tabular}{lrrr}
\hline
\multicolumn{1}{c}{} & \multicolumn{1}{c}{$J$} & \multicolumn{1}{c}{$H$}
 & \multicolumn{1}{c}{$K_s$} \\
\hline
Central wavelength [$\mu$m] & 1.25 & 1.63 & 2.14 \\
Bandwidth [$\mu$m] &  0.17 & 0.30 & 0.32 \\
\hline
\end{tabular}
\end{table}

In the previous paper (\citealt{ita2004}, hereafter referred to as Paper I), we cross-identified the OGLE data and the single-epoch SIRIUS $JHK$ data from the monitoring survey, and studied the pulsation properties and metallicity effects on period-$K$ magnitude relations by comparing the variable stars in the LMC and SMC. In this paper, we present the data of the variable stars in the Magellanic Clouds and discuss their infrared properties.

\section{DATA}
The total areas covered by the SIRIUS variable star survey are three and
one square degrees in the LMC and SMC, respectively. These areas are
divided into 27(9) regions of 20$^\prime$$\times$20$^\prime$ meshes in
the LMC(SMC). Tables~\ref{lmcarea} and \ref{smcarea} define the central
coordinates of them. Each of the 20$^\prime$$\times$20$^\prime$ regions
is further subdivided into nine sub regions due to the field of view of
the SIRIUS, which is about 7.7$^\prime$$\times$7.7$^\prime$.

\begin{table}
\caption{Survey regions in the Large Magellanic Cloud}
\label{lmcarea}
\centering
\begin{tabular}{crr}
\hline
\multicolumn{1}{c}{Name} & \multicolumn{1}{c}{Right Ascension} & \multicolumn{1}{c}{Declination} \\ 
\cline{2-3}
 & \multicolumn{2}{c}{Equinox: J2000} \\
\hline
LMC0507-6840 & 05$^\textrm{h}$ 07$^\textrm{m}$ 00.28$^\textrm{s}$ & -68$^{\circ}$ 40$^{\prime}$ 00.00$^{\prime\prime}$ \\ 
LMC0507-6900 & 05$^\textrm{h}$ 06$^\textrm{m}$ 50.29$^\textrm{s}$ & -69$^{\circ}$ 00$^{\prime}$ 00.00$^{\prime\prime}$ \\ 
LMC0507-6920 & 05$^\textrm{h}$ 06$^\textrm{m}$ 39.97$^\textrm{s}$ & -69$^{\circ}$ 20$^{\prime}$ 00.00$^{\prime\prime}$ \\ 
LMC0510-6920 & 05$^\textrm{h}$ 10$^\textrm{m}$ 26.65$^\textrm{s}$ & -69$^{\circ}$ 20$^{\prime}$ 00.00$^{\prime\prime}$ \\
LMC0510-6940 & 05$^\textrm{h}$ 10$^\textrm{m}$ 19.53$^\textrm{s}$ & -69$^{\circ}$ 40$^{\prime}$ 00.00$^{\prime\prime}$ \\
LMC0511-6900 & 05$^\textrm{h}$ 10$^\textrm{m}$ 33.53$^\textrm{s}$ & -69$^{\circ}$ 00$^{\prime}$ 00.00$^{\prime\prime}$ \\
LMC0514-6900 & 05$^\textrm{h}$ 14$^\textrm{m}$ 16.75$^\textrm{s}$ & -69$^{\circ}$ 00$^{\prime}$ 00.00$^{\prime\prime}$ \\
LMC0514-6920 & 05$^\textrm{h}$ 14$^\textrm{m}$ 13.31$^\textrm{s}$ & -69$^{\circ}$ 20$^{\prime}$ 00.00$^{\prime\prime}$ \\
LMC0514-6940 & 05$^\textrm{h}$ 14$^\textrm{m}$ 09.77$^\textrm{s}$ & -69$^{\circ}$ 40$^{\prime}$ 00.00$^{\prime\prime}$ \\
LMC0518-6900 & 05$^\textrm{h}$ 18$^\textrm{m}$ 00.00$^\textrm{s}$ & -69$^{\circ}$ 00$^{\prime}$ 00.00$^{\prime\prime}$ \\
LMC0518-6920 & 05$^\textrm{h}$ 18$^\textrm{m}$ 00.00$^\textrm{s}$ & -69$^{\circ}$ 20$^{\prime}$ 00.00$^{\prime\prime}$ \\
LMC0518-6940 & 05$^\textrm{h}$ 18$^\textrm{m}$ 00.00$^\textrm{s}$ & -69$^{\circ}$ 40$^{\prime}$ 00.00$^{\prime\prime}$ \\
LMC0522-6920 & 05$^\textrm{h}$ 21$^\textrm{m}$ 46.67$^\textrm{s}$ & -69$^{\circ}$ 20$^{\prime}$ 00.00$^{\prime\prime}$ \\
LMC0522-6940 & 05$^\textrm{h}$ 21$^\textrm{m}$ 50.23$^\textrm{s}$ & -69$^{\circ}$ 40$^{\prime}$ 00.00$^{\prime\prime}$ \\
LMC0522-7000 & 05$^\textrm{h}$ 21$^\textrm{m}$ 53.89$^\textrm{s}$ & -70$^{\circ}$ 00$^{\prime}$ 00.00$^{\prime\prime}$ \\
LMC0526-6920 & 05$^\textrm{h}$ 25$^\textrm{m}$ 33.35$^\textrm{s}$ & -69$^{\circ}$ 20$^{\prime}$ 00.00$^{\prime\prime}$ \\
LMC0526-6940 & 05$^\textrm{h}$ 25$^\textrm{m}$ 40.45$^\textrm{s}$ & -69$^{\circ}$ 40$^{\prime}$ 00.00$^{\prime\prime}$ \\
LMC0526-7000 & 05$^\textrm{h}$ 25$^\textrm{m}$ 47.81$^\textrm{s}$ & -70$^{\circ}$ 00$^{\prime}$ 00.00$^{\prime\prime}$ \\
LMC0529-6920 & 05$^\textrm{h}$ 29$^\textrm{m}$ 20.01$^\textrm{s}$ & -69$^{\circ}$ 20$^{\prime}$ 00.00$^{\prime\prime}$ \\
LMC0530-6940 & 05$^\textrm{h}$ 29$^\textrm{m}$ 30.67$^\textrm{s}$ & -69$^{\circ}$ 40$^{\prime}$ 00.00$^{\prime\prime}$ \\
LMC0530-7000 & 05$^\textrm{h}$ 29$^\textrm{m}$ 41.71$^\textrm{s}$ & -70$^{\circ}$ 00$^{\prime}$ 00.00$^{\prime\prime}$ \\
LMC0533-6940 & 05$^\textrm{h}$ 33$^\textrm{m}$ 20.91$^\textrm{s}$ & -69$^{\circ}$ 40$^{\prime}$ 00.00$^{\prime\prime}$ \\
LMC0534-7000 & 05$^\textrm{h}$ 33$^\textrm{m}$ 35.61$^\textrm{s}$ & -70$^{\circ}$ 00$^{\prime}$ 00.00$^{\prime\prime}$ \\
LMC0534-7020 & 05$^\textrm{h}$ 33$^\textrm{m}$ 50.83$^\textrm{s}$ & -70$^{\circ}$ 20$^{\prime}$ 00.00$^{\prime\prime}$ \\
LMC0537-6940 & 05$^\textrm{h}$ 37$^\textrm{m}$ 11.13$^\textrm{s}$ & -69$^{\circ}$ 40$^{\prime}$ 00.00$^{\prime\prime}$ \\
LMC0537-7000 & 05$^\textrm{h}$ 37$^\textrm{m}$ 29.52$^\textrm{s}$ & -70$^{\circ}$ 00$^{\prime}$ 00.00$^{\prime\prime}$ \\
LMC0538-7020 & 05$^\textrm{h}$ 37$^\textrm{m}$ 48.53$^\textrm{s}$ & -70$^{\circ}$ 20$^{\prime}$ 00.00$^{\prime\prime}$ \\
\hline
\end{tabular}
\end{table}

\begin{table}
\caption{Survey regions in the Small Magellanic Cloud}
\label{smcarea}
\centering
\begin{tabular}{crr}
\hline
\multicolumn{1}{c}{Name} & \multicolumn{1}{c}{Right Ascension} & \multicolumn{1}{c}{Declination} \\ 
\cline{2-3}
 & \multicolumn{2}{c}{Equinox: J2000} \\
\hline
SMC0050-7250 & 00$^\textrm{h}$ 50$^\textrm{m}$ 28.95$^\textrm{s}$ & -72$^{\circ}$ 50$^{\prime}$ 00.00$^{\prime\prime}$ \\
SMC0050-7310 & 00$^\textrm{h}$ 50$^\textrm{m}$ 23.74$^\textrm{s}$ & -73$^{\circ}$ 10$^{\prime}$ 00.00$^{\prime\prime}$ \\
SMC0051-7230 & 00$^\textrm{h}$ 50$^\textrm{m}$ 33.95$^\textrm{s}$ & -72$^{\circ}$ 30$^{\prime}$ 00.00$^{\prime\prime}$ \\
SMC0055-7230 & 00$^\textrm{h}$ 55$^\textrm{m}$ 00.00$^\textrm{s}$ & -72$^{\circ}$ 30$^{\prime}$ 00.00$^{\prime\prime}$ \\
SMC0055-7250 & 00$^\textrm{h}$ 55$^\textrm{m}$ 00.00$^\textrm{s}$ & -72$^{\circ}$ 50$^{\prime}$ 00.00$^{\prime\prime}$ \\
SMC0055-7310 & 00$^\textrm{h}$ 55$^\textrm{m}$ 00.00$^\textrm{s}$ & -73$^{\circ}$ 10$^{\prime}$ 00.00$^{\prime\prime}$ \\
SMC0059-7230 & 00$^\textrm{h}$ 59$^\textrm{m}$ 26.04$^\textrm{s}$ & -72$^{\circ}$ 30$^{\prime}$ 00.00$^{\prime\prime}$ \\
SMC0100-7250 & 00$^\textrm{h}$ 59$^\textrm{m}$ 31.04$^\textrm{s}$ & -72$^{\circ}$ 50$^{\prime}$ 00.00$^{\prime\prime}$ \\
SMC0100-7310 & 00$^\textrm{h}$ 59$^\textrm{m}$ 36.25$^\textrm{s}$ & -73$^{\circ}$ 10$^{\prime}$ 00.00$^{\prime\prime}$ \\
\hline
\end{tabular}
\end{table}

We cross-identified OGLE variables in the Magellanic Clouds (\citealt{zebrun2001}) with the SIRIUS NIR sources that have been detected at least two of the three ($JHK$) wave bands, and then determined their pulsation periods by using the Phase Dispersion Minimization (PDM) technique \citep{stelling}. Details of the cross-identifications and period findings are described in Paper I. We did not analyze multi-periodic stars (e.g., \citealt{bedding}) and/or variable stars that show too complex light curves in the phase space (see the bottom panel of figures~\ref{largetheta} and ~\ref{smalltheta}) to find a predominant period. We made two catalogs presenting data of variable stars in the LMC and SMC. They contain the following information: OGLE name, pulsation period in days, statistical parameter $\theta$ calculated by the PDM, pulsation amplitude ($\Delta I = I_{\textrm{max}} - I_{\textrm{min}}$), intensity mean $I$ magnitude $<I>$, $JHK$ magnitudes and positional difference $r$ between the SIRIUS coordinate and the OGLE coordinate in arcsec. The $\theta$ is a measure of the regularity of the light variation, being near 0 for the regular variation and near 1 for the irregular variation.

Table~\ref{sample} is a sample that shows the first five records of the LMC catalog. The full version of the data, including the SMC catalog is available in the on-line version of this paper. In cases the $JHK$ measurements are saturated or below the detection limit, the corresponding columns contain the value 99.999. The $JHK$ magnitudes in the catalog are referred to the Las Campanas Observatory (LCO) system ones based on observations of a few dozen stars from \citet{persson} and are not corrected for the interstellar absorption. We used the following colour equations
\begin{gather}
J_{\textrm{LCO}} = J + 
\left(
\begin{array}{r}
-0.007\pm0.002 \\
-0.005\pm0.002 \\
-0.003\pm0.002 \\
\end{array}
\right) +
\\
%\left(
%0.016\pm0.011~~0.009\pm0.005~~0.000\pm0.000 
%\right)
\left(
\begin{array}{r}
0.016\pm0.011(J-H)\\
0.009\pm0.005(J-K)\\
0.000\pm0.000(H-K)\\
\end{array}
\right) \nonumber
\end{gather}
\begin{gather}
H_{\textrm{LCO}} = H +
\left(
\begin{array}{r}
0.001\pm0.002 \\
-0.002\pm0.002 \\
0.000\pm0.002 \\
\end{array}
\right) +
\\
%\left(
%-0.001\pm0.000~~0.004\pm0.000~~0.004\pm0.000
%\right)
\left(
\begin{array}{r}
-0.001\pm0.000(J-H)\\
0.004\pm0.000(J-K)\\
0.004\pm0.000(H-K)\\
\end{array}
\right) \nonumber
\end{gather}
\begin{gather}
K_{\textrm{LCO}} = K +
\left(
\begin{array}{r}
-0.001\pm0.004 \\
-0.001\pm0.004 \\
0.005\pm0.004 \\
\end{array}
\right) +
\\
%\left(
%-0.002\pm0.000~~-0.002\pm0.000~~-0.041\pm0.004 
%\right)
\left(
\begin{array}{c}
-0.002\pm0.000(J-H)\\
-0.002\pm0.000(J-K)\\
-0.041\pm0.004(H-K)\\
\end{array}
\right) \nonumber
\end{gather}
to transform the IRSF instrumental magnitudes to the LCO system based
ones. The $J-K$ ones were usually used, but $J-H$($H-K$) ones were employed if the $K$($J$) magnitudes were unavailable. Note that the SIRIUS measures the $K_s$ magnitudes and they were converted to the LCO $K$ magnitudes.

To enhance the image quality, we stacked the top 10 best seeing
images\footnote{These 10 images were taken on different days and hence
under the different conditions, and a single image comprises 10 dithered
5 sec exposures.} (typically the seeing ranges from about 0.9$^{\prime\prime}$ to
1.2$^{\prime\prime}$ at $K_s$) that we have been collecting in the ongoing SIRIUS
monitoring survey. We used the ISIS.V2.1 package (\citealt{alard1998};
\citealt{alard2000}) to build a composite image that comprises 100
dithered 5 sec exposures. Photometry was performed on the composite
image with DoPHOT (\citealt*{schechter}), and then, the measured
magnitudes were calibrated with the single-epoch photometry data by matching
the magnitudes of the non-variable stars. Eventually, the photometry has
a signal-to-noise ratio (S/N) of about 10 at 17.40, 17.38 and 16.63 mag
in $J$, $H$ and $K$, respectively. Also, traditional experiments of adding
artificial stars to the composite images revealed that the 90\%
completeness (detection probability) limits at $J$, $H$ and $K$ are about 17.25, 17.22 and 16.45 mag, respectively (\citealt{yitaphd}).

\begin{table*}
\caption{The first 5 records in the LMC catalog. This is a sample of the full version, which is available in the electronic version of Monthly Notices.}
\label{sample}
\centering
\begin{tabular}{crrrrrrrr}
\hline
\multicolumn{1}{c}{Name} & \multicolumn{1}{c}{Period} & \multicolumn{1}{c}{$\theta$} & \multicolumn{1}{c}{$\Delta I$} & \multicolumn{1}{c}{$<I>$} & \multicolumn{1}{c}{$J$} & \multicolumn{1}{c}{$H$} & \multicolumn{1}{c}{$K$} & \multicolumn{1}{c}{r} \\
\cline{4-8}
\multicolumn{1}{c}{} & \multicolumn{1}{c}{[days]} & \multicolumn{1}{c}{} & \multicolumn{5}{c}{[mag]} & \multicolumn{1}{c}{[arcsec]} \\
\hline
OGLE050437.94-692217.2 &    0.460 & 0.178 &  1.025 & 19.250 & 99.999 & 16.056 & 16.030 & 2.260 \\ 
OGLE050438.43-691527.2 &   49.892 & 0.837 &  0.089 & 13.951 & 13.559 & 10.786 & 11.337 & 0.306 \\ 
OGLE050440.26-693039.7 &   42.783 & 0.866 &  0.160 & 14.189 & 12.786 & 11.880 & 11.599 & 0.466 \\ 
OGLE050440.29-692708.3 &   39.200 & 0.661 &  0.096 & 14.734 & 13.423 & 12.617 & 12.333 & 0.465 \\ 
OGLE050440.76-692416.0 &   96.365 & 0.358 &  0.317 & 14.060 & 12.294 & 11.491 & 11.063 & 0.574 \\ 
\hline
\end{tabular}
\end{table*}

\section{Discussion}
In the following discussions, we use the $JHK$ magnitudes that are corrected for the interstellar absorption based on the relations in \citet{koornneef}, assuming $R =$ 3.2. We adopted ($A_J$, $A_H$, $A_K$) $=$ (0.129, 0.084, 0.040) and (0.082, 0.053, 0.025), corresponding to the total mean reddening of $E_{B-V} =$ 0.137 and 0.087 as derived by \citet{udalski1999} for the OGLE's observing fields in the LMC and SMC, respectively. The reddening of the Magellanic Clouds is still somewhat controversial (e.g., \citealt{westerlund}), and it is certain that this type of statistical average approach can not account for individual corrections. However, the adopted reddening should not affect any of the conclusions of this paper. The coefficients we calculated in equations~\ref{pccepheidlmc} to \ref{pcsmc} depend on reddening, but the effects should be small.

\begin{figure}
\centering
\includegraphics[angle=0,scale=0.455]{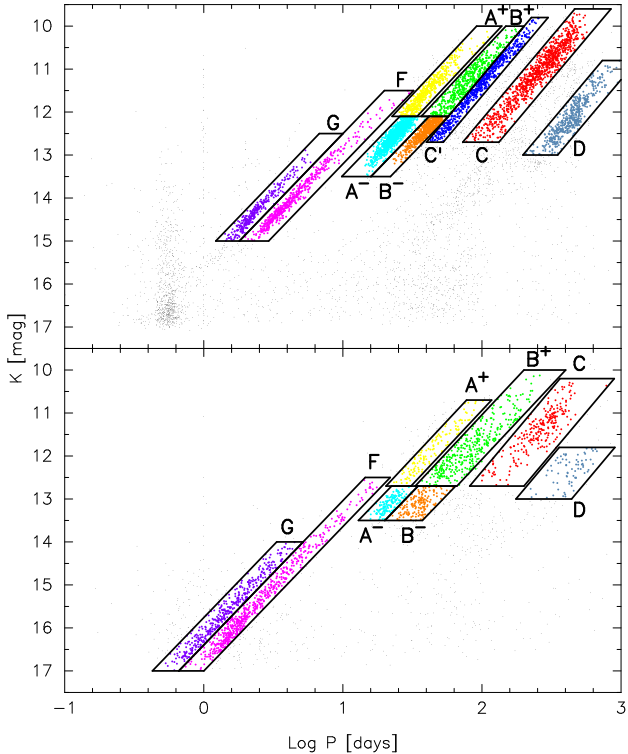}
\caption{Variable stars in the LMC (upper panel) and SMC (lower panel) are classified into several groups according to their location on the period-$K$ magnitude plane.}
\label{plk}
\end{figure}

\begin{table}
\caption{Colour mapping table}
\label{colourmap}
\centering
\begin{tabular}{lll}
\hline
\multicolumn{1}{l}{Label} & \multicolumn{1}{l}{Colour} & \multicolumn{1}{c}{Population} \\
\hline
$A^-$ & Cyan & RGB variables \& metal poor and old AGB variables \\
$A^+$ & Yellow & less regularly pulsating AGB variables \\
$B^-$ & Orange & RGB variables \& metal poor and old AGB variables \\
$B^+$ & Green & less regularly pulsating AGB variables \\
$C^\prime$ & Blue & Mira variables pulsating in the 1st overtone mode\\
$C$ & Red & Mira variables pulsating in the fundamental mode \\
$D$ & Steel-blue & Some obscured variables \& Unknown variables \\
$F$ & Magenta & Cepheid variables pulsating in the fundamental mode \\
$G$ & Purple & Cepheid variables pulsating in the 1st overtone mode \\
\hline
\end{tabular}
\end{table}

\subsection{Classification of the variable stars on the period-$K$
  magnitude plane}
\citet{wood1999} discovered that variable red giants in the LMC form
parallel sequences in the period-$K$ magnitude plane and suggested that
these sequences can be interpreted as the differences in the pulsation
modes (\citealt{wood2000}).

In Paper I, we classified variable stars into nine (LMC) and eight (SMC) prominent groups based on their locations on the period-$K$ magnitude plane. Refer to Paper I for the details of the grouping. For the easier reading and completeness of this paper, we show the period-$K$ magnitude diagram with classification boxes in figure~\ref{plk}, and also for conciseness, we tabulated the colour mapping chart in table~\ref{colourmap}. Stars outside the classification boxes are represented by black tiny dots. Throughout this paper, we use the same labels and colours for each group. In the LMC, there is a clump of variables around $\log P \lesssim$ $-$0.1 and $K \gtrsim$ 15.0. This is a group of RR Lyrae variables, which we do not analyze in this paper.

\begin{figure}
\centering
\includegraphics[angle=0,scale=0.455]{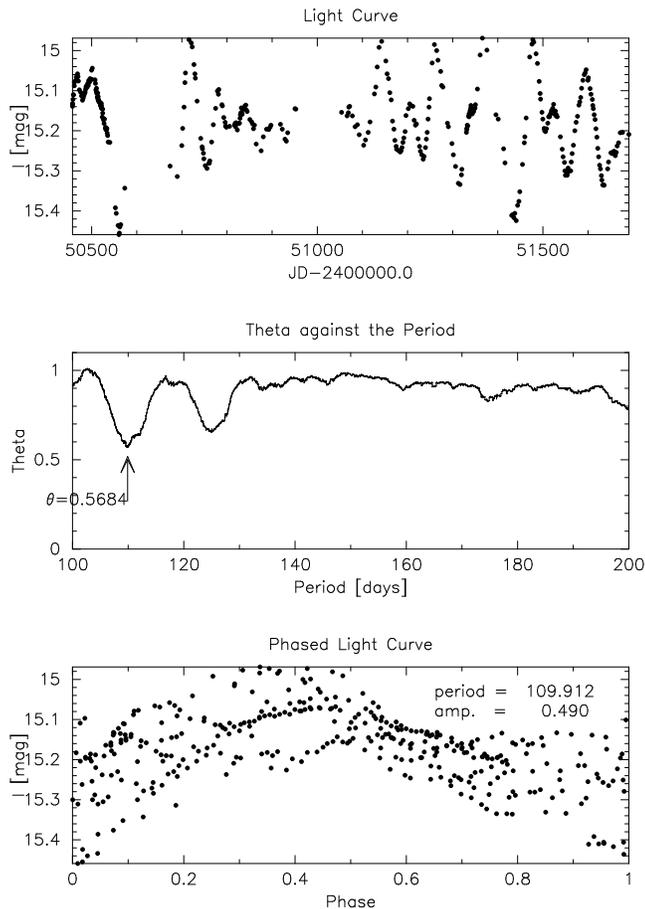}
\caption{The light curve, theta spectrum and phased light curve (from top to the bottom) of the star OGLE051707.45-692855.3.}
\label{largetheta}
\end{figure}

\begin{figure}
\centering
\includegraphics[angle=0,scale=0.455]{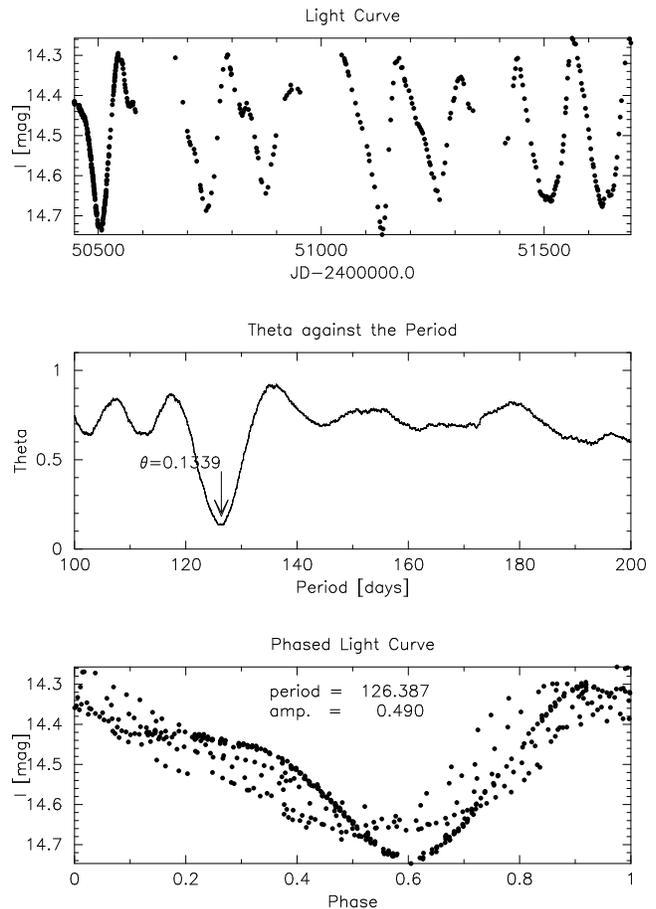}
\caption{The same as figure~\ref{largetheta}, but for the star
 OGLE052353.10-694718.4 to show a similar star with different value of $\theta$.}
\label{smalltheta}
\end{figure}

Roughly speaking, five types of variable stars constitute the majority of the catalog; Cepheids, Miras, Semi-regulars, Irregular variables and Eclipsing binaries. However, telling Semi-regulars from Mira variables is rather difficult and there are not necessarily deep physical criteria. Several authors (e.g., \citealt{alard}; \citealt{cioni2001}) use a simple criterion that is based on a certain limiting pulsation amplitude to separate Semi-regulars and Miras. Meanwhile, \citet[2004]{kiss} showed that there is a good correlation between the amplitude and mode of pulsation. In this paper, we do not separate Semi-regulars and Miras explicitly. Instead, we distinguish ``regularly pulsating variables, ($\theta \leq$ 0.55)'' from ``less regularly pulsating variables, ($\theta > 0.55$)'' relying on the parameter ``$\theta$''. In figures~\ref{largetheta} and \ref{smalltheta}, we show the representative light curves of two variable stars on sequence $C$, having similar periods ($\sim$109 and $\sim$126 days) and the same amplitudes ($\Delta I\sim$0.490 mag) but different $\theta$s.

\subsection{Colour-magnitude and colour-colour diagram}
\begin{figure}
\centering
\includegraphics[angle=0,scale=0.455]{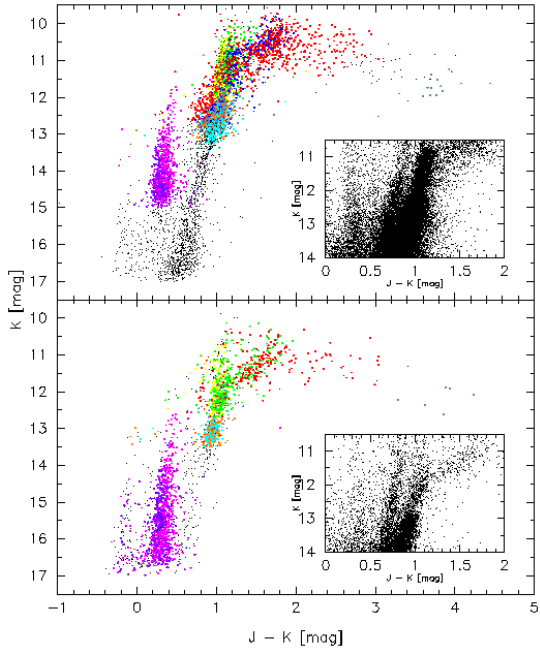}
\caption{NIR colour-magnitude diagram of OGLE variables in the LMC (upper panel) and the SMC (lower panel). The inset shows the plot of the all SIRIUS data (including non-variables) around the $K$-band luminosity of the TRGB.}
\label{colmag}
\end{figure}

\begin{figure}
\centering
\includegraphics[angle=0,scale=0.455]{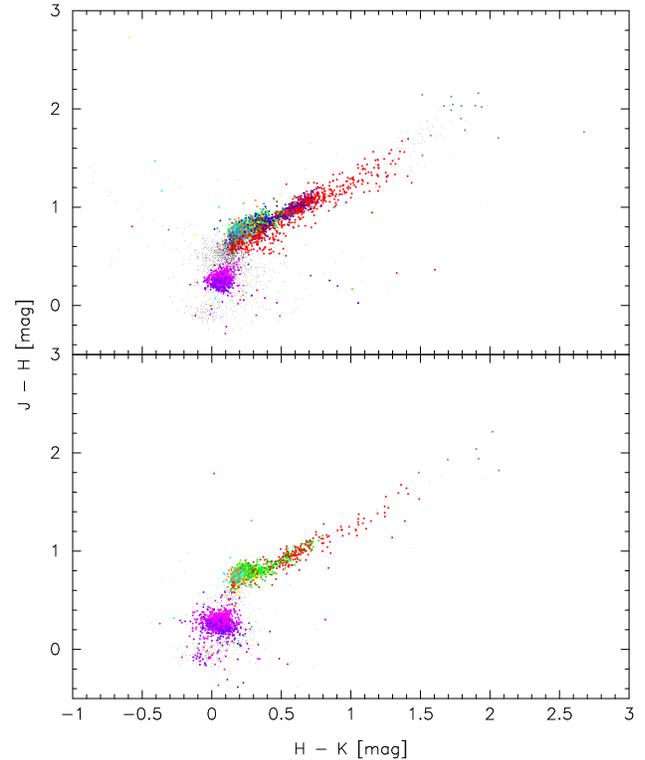}
\caption{NIR colour-colour diagram of OGLE variables in the LMC (upper panel) and the SMC (lower panel).}
\label{colcol}
\end{figure}

The NIR data from the SIRIUS survey enabled us to study the infrared properties of the OGLE variables. Figure~\ref{colmag} and \ref{colcol} show the NIR colour-magnitude and colour-colour diagram of the OGLE variables. In the inset of the figure~\ref{colmag}, we show the colour-magnitude diagram of the SIRIUS data (including non-variables), closing-up around the tip of the first giant branch (TRGB). By comparing the insets and the main diagrams, it is clear that the stars of group $A^-$ and $B^-$ accumulate at the exact location of the TRGB (\citealt{ita}; \citealt[2004]{kiss}; \citealt{ita2004}).

Most of the stars from group $D$ are found at the place a bit brighter
than the TRGB. \citet{wood1999} identified the sequence $D$ with
suspected binaries, and \citet{wood2004} discussed the possible
explanations for the $D$ variables. On the other hand, some of the $D$ variables are
very red ($J - H > 1.8$, $H - K > 1.5$ and $J - K > 3.2$) and relatively
bright (11.0 $\lesssim K \lesssim$ 12.5), properties associated with
carbon-rich stars and dusty AGB stars (\citealt{nikolaev}). In the
present sample, 14 out of 471 (6 out of 71) $D$ stars in the LMC (SMC)
have these properties. A possible explanation for these red stars in the
group $D$ is that, these stars could in fact be very dusty AGB stars that are subjected to severe circumstellar extinctions, even in the $K$ band. \citet{whitelock2003} studied obscured variables in the LMC and showed that the $K$ magnitudes of obscured stars cover a wide range at a given pulsation period. For instance, their data indicates that the difference can be more than two magnitudes in the case of the carbon-rich stars. It is interesting to note that their red stars (those with long periods and $K$ magnitudes fainter than $\sim$11 mag) clearly locate the brighter part of sequence $D$ on the period-$K$ magnitude plane. In addition, \citet{kiss} suggested that there is a group of stars located below the Mira sequence (sequence $C$) and identified them with dusty AGB stars.

\begin{figure}
\centering
\includegraphics[angle=0,scale=0.455]{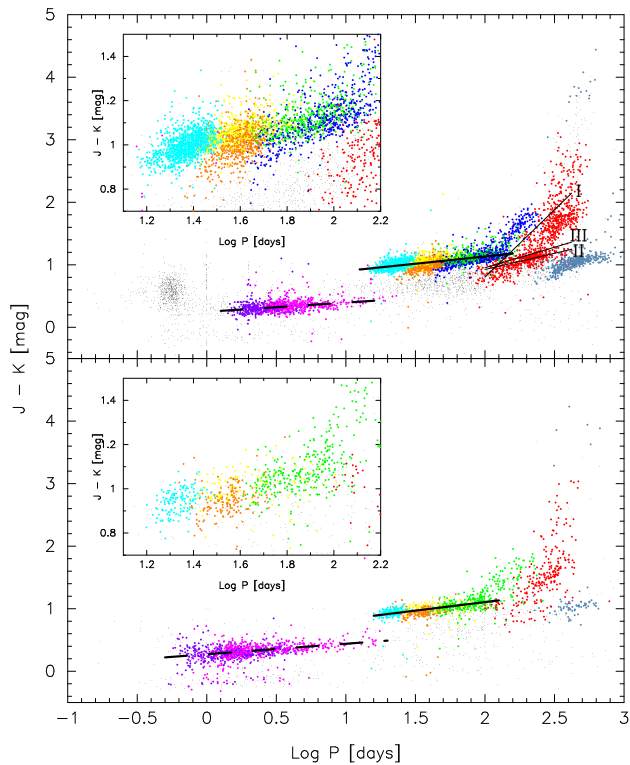}
\caption{Period-colour diagram of OGLE variables in the LMC (upper panel) and the SMC (lower panel). The dashed and thick solid lines indicate the least-square fits of linear relation to each distinct sequence. The thin solid lines labeled by I to III in the upper panel are the period-$J - K$ colour relations for Mira variables obtained by \citet{feast1989} and \citet{whitelock2000} (see text). The insets show the close-ups of less regularly pulsating variables.}
\label{periodcolor}
\end{figure}

\subsection{Period-colour diagram}
Figure~\ref{periodcolor} shows the $J - K$ colours of OGLE variables in the Magellanic clouds as a function of their periods. The insets are the close-ups for the sake of clarity. A general trend that the colours get redder with increasing period can be seen. 

The dashed lines are the least-square fit of a linear relation to the Cepheid variables ($F$ and $G$ stars), which yields the period-$J - K$ colour relations of
\begin{equation}
J - K \textrm{[mag]} = 0.147(\pm0.010) \log P \textrm{[days]} + 0.248(\pm0.006)
\label{pccepheidlmc}
\end{equation}
for the LMC ($\sigma$ = 0.062) and
\begin{equation}
J - K \textrm{[mag]} = 0.167(\pm0.009) \log P \textrm{[days]} + 0.272(\pm0.004)
\end{equation}
for the SMC ($\sigma$ = 0.074), respectively. The quoted errors are the 1$\sigma$ errors of each coefficient. Although the fundamental (group $F$) and first overtone (group $G$) Cepheids are clearly separated in the period-$K$ magnitude plane, we don't see any significant difference between them in this diagram. This means, if we compare the fundamental and first overtone Cepheids with the same pulsation periods, the stellar radii of first overtone Cepheids should be larger than those of fundamental Cepheids because their stellar temperatures are nearly the same but the first overtone Cepheids are brighter than the fundamental ones. Theories predict exactly the same conclusion (e.g., \citealt{bono}), and if the stellar radius of the Cepheid variables could be measured accurately, it will help to identify the pulsation modes.

The less regularly pulsating red giants (group $A^{\pm}$ and $B^{\pm}$) also follow a tight sequence on the period-$J - K$ colour plane. The thick solid lines in the figure indicate the least-square fit of a linear relation to the stars from the four groups, whose $J - K$ colours and periods are within the ranges, 0.7 $< J - K <$ 1.3 and 1.2 $< \log P <$ 2.1. Their period-$J - K$ colour relations are also calculated and they are,
\begin{equation}
J - K \textrm{[mag]} = 0.231(\pm0.004) \log P \textrm{[days]} + 0.671(\pm0.007).
\label{pclmc}
\end{equation}
for the LMC ($\sigma$ = 0.058) and
\begin{equation}
J - K \textrm{[mag]} = 0.272(\pm0.011) \log P \textrm{[days]} + 0.560(\pm0.019).
\label{pcsmc}
\end{equation}
for the SMC ($\sigma$ = 0.068), respectively.

It is obvious that most of the $C$ and $C^\prime$ stars in the LMC, and
also some $B^+$ stars and most of $C$ stars in the SMC do not follow the
extension of these relations. In Paper I, we confirmed the conclusion of
\citet{wood1996} and suggested that stars on sequences $C$ and
$C^\prime$ are Mira variables pulsating in the fundamental and first
overtone mode respectively, by comparing their model calculations with our observational data. Figure~\ref{periodcolor} clearly shows that $C$ and $C^\prime$ stars behave very similarly on the period-$J - K$ colour plane. This corroborates the idea that the $C$ and $C^\prime$ stars are the same type of stars (i.e., Miras), but only different in the pulsation modes. Also, it is likely that the ``some $B^+$ stars'' in the SMC are the counterparts of the $C^\prime$ stars in the LMC.

In earlier work, \citet{feast1989}, \citet{glass1995} and \citet*{whitelock2000} obtained period-$J - K$ colour relations of Mira variables in the LMC, Sgr I Baade window of the Galactic Bulge and solar neighborhood, respectively. These relations are all based on the SAAO system. The atmospheres of Mira variables are rather complex, making their quantitative comparisons difficult between different filter systems. Thus here we make only qualitative discussions. Just to get a rough idea, we indicated in the upper panel of figure~\ref{periodcolor} the period-$J - K$ colour relations obtained by \citet{feast1989} for carbon- (labeled I) and oxygen-rich (labeled II) Miras and by \citet{whitelock2000} (labeled III) after referring to the LCO system by assuming $(J - K)_{\textrm{LCO}} = 0.928 (J - K)_{\textrm{SAAO}} - 0.004$ \citep{carpenter2001}. Of course we can not tell exactly which of our stars are carbon- or oxygen-rich, but it is statistically fair to say that stars with $J - K$ colours redder than 1.4 are primarily carbon-rich stars (\citealt{nikolaev}). In this point of view, we confirmed the result of \citet{feast1989} that oxygen- and carbon-rich Miras follow different period-$J - K$ colour relations.

Now that we see the $C$ and $C^\prime$ stars separate on the period-$J - K$ colour plane, it will be interesting to search for the Galactic counterpart of the first overtone Miras ($C^\prime$ stars) by using this tool. Then we can redetermine their distances using their exclusive period-$K$ magnitude relation, which we obtained in Paper I. It will help to understand the differences between the fundamental and first overtone Miras through more detailed studies that are technically difficult in the Magellanic Clouds (i.e., spectroscopic work and mid-infrared observations etc.). Because not only $C$ and $C^\prime$ stars but also the other variables separate on the period-$J - K$ colour plane, the same argument is true for them, such as $D$ stars. 

However, before applying this tool to different environments, we should know whether it depends on chemical abundance. \citet*{wray2004} showed that small amplitude red giant variables in the Galactic Bar can be separated into two groups, and they follow different period-colour relations. Also, they suggest that their $A$ and $B$ stars correspond to type $A^-$ and $B^-$ variables, respectively of Paper I. The insets of the figure~\ref{periodcolor} clearly confirm their result, showing that $A^-$ and $B^-$ are well separated on the period-$J - K$ color plane. This infers that the period-$J - K$ relations we derived in equations~\ref{pclmc} and \ref{pcsmc} are merely the general trends of the variable red giants, and in fact there are exclusive period-colour relations for each population. \citet{rejkuba2003} studied infrared properties of long period variables in Cen A. Although they did not mention it, it is likely that there are two groups in their period-colour diagrams (see their figure 10). We suggest that their short-period ($\log P\sim2.45$) and long-period ($\log P\sim2.65$) groups could correspond to type $C^\prime$ and $C$ stars in the LMC, respectively. These facts strengthen the idea that, period-colour diagram is a good and universal tool, at least to tell the type of variable stars within a system.

\begin{figure}
\centering
\includegraphics[angle=0,scale=0.455]{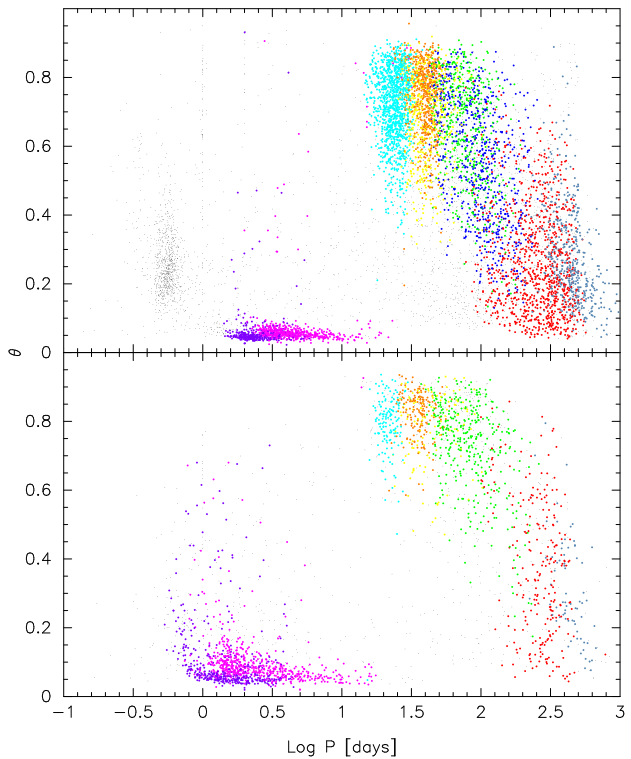}
\caption{Period-$\theta$ diagram of OGLE variables in the LMC (upper panel) and the SMC (lower panel).}
\label{p-t}
\end{figure}

\citet*{lebzelter} studied the AGAPEROS/DENIS variables in the LMC and showed that so-called Semi-regular variables follow a period-colour relation but the ``regular'' variables do not (see their figure 12). To see the regularity of the light variation quantitatively, we show the relationship between the statistic parameter $\theta$ and period in figure~\ref{p-t}. The figure clearly confirmed the Lebzelter's suggestion and shows that stars that follow period-$J - K$ relation ($A^\pm$ and $B^\pm$ stars with $\log P \lesssim$ 2.1) are less regularly pulsating, and those that don't ($B^+$ stars with $\log P \gtrsim 2.1$, $C^\prime$ and $C$ stars) are regularly pulsating.

\subsection{Period-amplitude ($\Delta I$) and colour-amplitude diagram}
\begin{figure}
\centering
\includegraphics[angle=0,scale=0.455]{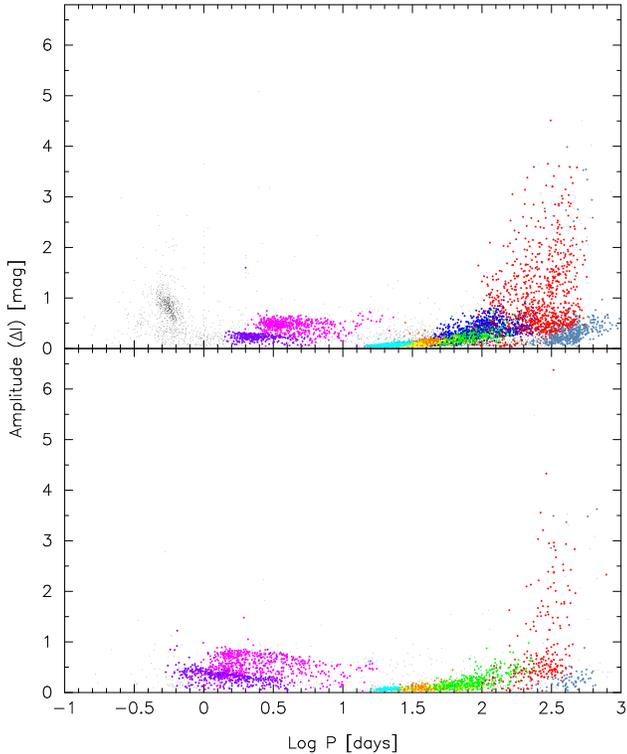}
\caption{Period-amplitude($\Delta I$) diagram of OGLE variables in the LMC (upper panel) and the SMC (lower panel).}
\label{p-a}
\end{figure}

Figure~\ref{p-a} is a plot to show the relationship between the pulsation period and amplitude ($\Delta I$). The pulsation amplitude is determined from the OGLE's $I$ band data, given as $\Delta I = I_{\textrm{max}} - I_{\textrm{min}}$. Note that all of the original OGLE light curves were carefully eye-inspected, and we calculated the $\Delta I$ after eliminating the obvious spurious data points.

The figure shows that the fundamental Cepheids show larger amplitude variations than the first overtone Cepheids. Also, the upper envelope of Cepheid amplitude goes down with increasing period. This trend contrasts markedly with that of red giants, because their pulsation amplitudes generally go up with increasing period.

Variables stars with $K$ magnitudes below the TRGB ($A^-$ and $B^-$ stars) pulsate with very small amplitude. With a few exceptions, most of the stars from $D$ group also pulsate with relatively small amplitude. The large amplitude pulsation ($\Delta I \gtrsim$ 0.9) seems to occur only among the long-period ($\log P \gtrsim$ 2.3) variables. If one had to separate Semi-regular from Mira variables based on pulsation amplitudes, the threshold would be about $\Delta I \approx$ 0.9, because one can see a discernible gap around there.

\begin{figure}
\centering
\includegraphics[angle=0,scale=0.455]{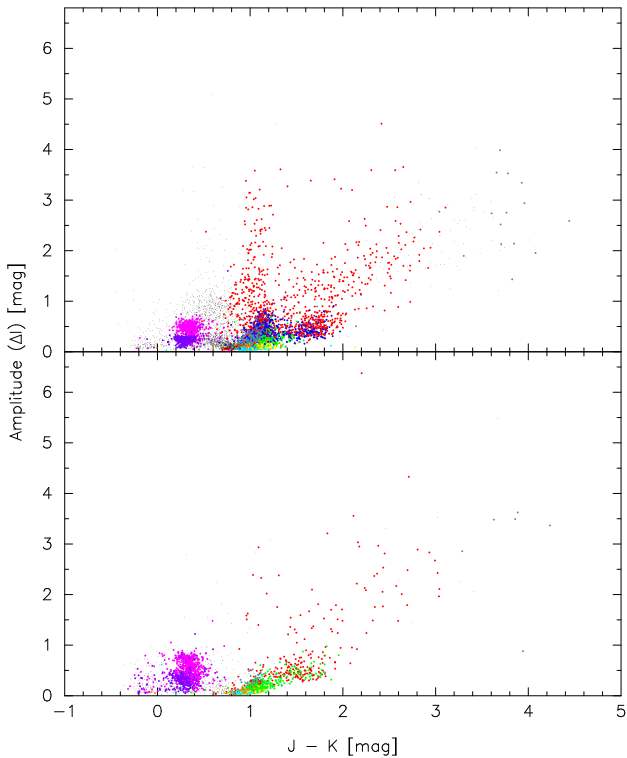}
\caption{Colour-amplitude($\Delta I$) diagram of OGLE variables in the LMC (upper panel) and the SMC (lower panel).}
\label{c-a}
\end{figure}

Figure~\ref{c-a} shows the relationship between the $J - K$ colour and pulsation amplitude ($\Delta I$). The variable red giants show the clear trend that the amplitude gets larger with increasing colour. According to \citet{nikolaev}, obscured carbon-rich AGB stars have the $J - K$ colours redder than about 2.0 mag. Almost all of such dusty carbon-rich stars in the Magellanic Clouds have large pulsation amplitudes. Some of the relatively blue stars (0.8 $\lesssim J - K \lesssim$ 1.2) on the sequence $C$ also show large amplitude pulsation. Judging from their $J - K$ colour and being on the sequence $C$, they are likely to be oxygen-rich Mira variables. It is remarkable that the carbon-rich Miras tend to have greater $I$-band amplitudes at redder $J - K$ colour, but the amplitudes of oxygen-rich Miras are almost independent of the colour. This might reflect the difference in the molecular opacities at $I$-band between the two groups.

\subsection{Similarities and differences between the LMC and SMC samples}
In a large sense, variable stars in the Magellanic Clouds behave similarly in the figures we discussed so far. However, on closer view, there are differences between variable stars in the two galaxies. In the colour-amplitude diagram (figure~\ref{c-a}), we see smaller number of oxygen-rich large amplitude Mira variables in the SMC (counterpart of LMC Miras with $J - K$ colours bluer than 1.4) compared to the LMC. This is probably because the difference in average chemical abundance and/or in age between the two galaxies (e.g., \citealt{cioni2003}; \citealt{mouhcine}).

\section{Summary}
We presented the data of 8,852 and 2,927 variable stars in the Large and Small Magellanic Clouds. Based on these data, we discussed the infrared properties of the variable stars. We showed that period-$J - K$ colour diagram is a good tool to tell the pulsation modes of the Mira variables. A follow-up campaign searching for the Galactic counterpart of the first overtone mode Miras would tell us the difference between them and fundamental mode Miras.

\section*{Acknowledgments}
We thank the referee for the constructive and helpful comments that improved this paper. We are grateful to people involved in OGLE project for making their data so easy to access and use. We would like to thank Dr. Michael Feast and Dr. Ian Glass for the valuable and helpful comments on the first version of the manuscript. We also thank Dr. Bohdan Paczy\'{n}ski and Dr. Tim Bedding for the kind comments. This research is supported in part by the Grant-in-Aid for Scientific Research (C) No. 12640234 and Grant-in-Aids for Scientific on Priority Area (A) No. 12021202 and 13011202 from the Ministry of Education, Science, Sports and Culture of Japan. The IRSF/SIRIUS project was initiated and supported by Nagoya University, National Astronomical Observatory of Japan and University of Tokyo in collaboration with South African Astronomical Observatory under a financial support of Grant-in-Aid for Scientific Research on Priority Area (A) No. 10147207 of the Ministry of Education, Culture, Sports, Science, and Technology of Japan.

\label{lastpage}


\begin{thebibliography}{}

  \bibitem[\protect\citeauthoryear{Alard \& Lupton}{1998}]{alard1998} Alard C., Lupton R. H., 1998,ApJ, 503, 325

  \bibitem[\protect\citeauthoryear{Alard}{2000}]{alard2000} Alard C., 2000, A\&AS, 144, 363

  \bibitem[\protect\citeauthoryear{Alard et al.}{2001}]{alard} Alard C. et al., 2001, ApJ, 552, 289

%  \bibitem[\protect\citeauthoryear{Antonello, Fugazza \& Mantegazza}{Antonello et al.}{2002}]{antonello} Antonello E., Fugazza D., Mantegazza L., 2002, A\&A, 388, 477

  \bibitem[\protect\citeauthoryear{Bedding et al.}{1998}]{bedding} Bedding T. R., Zijlstra Albert A., Jones A., Foster G., 1998, MNRAS, 301, 1073

  \bibitem[\protect\citeauthoryear{Bono et al.}{2001}]{bono} Bono G., Gieren W. P., Marconi M., Fouqu\'e, 2001, ApJ, 552, L141

  \bibitem[\protect\citeauthoryear{Carpenter}{2001}]{carpenter2001} Carpenter J. M., 2001, AJ, 121, 2851

  \bibitem[\protect\citeauthoryear{Cioni et al.}{2001}]{cioni2001} Cioni M.-R. L., Marquette J.-B., Loup C., Azzopardi M., Habing H. J., Lasserre T., Lesquoy E., 2001, A\&A, 377, 945

  \bibitem[\protect\citeauthoryear{Cioni \& Habing}{2003}]{cioni2003} Cioni M.-R. L., Habing H. J., 2003, A\&A, 402, 133

  \bibitem[\protect\citeauthoryear{Feast et al.}{1989}]{feast1989} Feast M. W., Glass I. S., Whitelock P. A., Catchpole R. M., 1989, MNRAS, 241, 375

  \bibitem[\protect\citeauthoryear{Glass et al.}{1995}]{glass1995} Glass I. S., Whitelock P. A., Catchpole R. M., Feast M. W., 1995, MNRAS, 273, 383

  \bibitem[\protect\citeauthoryear{Ita et al.}{2002}]{ita} Ita Y. et al., 2002, MNRAS, 337, L31

  \bibitem[\protect\citeauthoryear{Ita et al.}{2004a}]{ita2004} Ita Y. et al., 2004a, MNRAS, 347, 720 (Paper I)

  \bibitem[\protect\citeauthoryear{Ita}{2004b}]{yitaphd} Ita Y., 2004b, PhD thesis, Univ. Tokyo

  \bibitem[\protect\citeauthoryear{Kiss \& Bedding}{2003}]{kiss} Kiss L. L., Bedding T. R., 2003, MNRAS, 343, L79

  \bibitem[\protect\citeauthoryear{Kiss \& Bedding}{2004}]{kiss2004} Kiss L. L., Bedding T. R., 2004, MNRAS, 347, L83

  \bibitem[\protect\citeauthoryear{Koornneef}{1982}]{koornneef} Koornneef J., 1982, A\&A, 107, 247

  \bibitem[\protect\citeauthoryear{Lebzelter, Schultheis \& Melchior}{Lebzelter et al.}{2002}]{lebzelter} Lebzelter T., Schultheis M., Melchior A. L., 2002, A\&A, 393, 573

  \bibitem[\protect\citeauthoryear{Mouhcine \& Lan\c{c}on}{2003}]{mouhcine} Mouhcine M., Lan\c{c}on A., 2003, MNRAS, 338, 572

  \bibitem[\protect\citeauthoryear{Nagashima et al.}{1999}]{nagashima} Nagashima C. et al., 1999, in Nakamoto T., ed., Star formation 1999, Nobeyama Radio Observatory, p. 397

  \bibitem[\protect\citeauthoryear{Nagayama et al.}{2002}]{nagayama} Nagayama T. et al., 2002, Proc. SPIE, 4841, 459

  \bibitem[\protect\citeauthoryear{Nikolaev \& Weinberg}{2000}]{nikolaev} Nikolaev S., Weinberg M. D., 2000, ApJ, 542, 804

  \bibitem[\protect\citeauthoryear{Persson et al.}{1998}]{persson} Persson S. E., Murphy D. C., Krzeminski W., Roth M., Rieke M. J., 1998, AJ, 116, 2475

  \bibitem[\protect\citeauthoryear{Rejkuba et al.}{2003}]{rejkuba2003} Rejkuba M., Minniti D., Silva D. R., Bedding T. R., 2003, A\&A, 411, 351

  \bibitem[\protect\citeauthoryear{Schechter, Mateo \& Saha}{Schechter et al.}{1993}]{schechter} Schechter P. L., Mateo M., Saha A., 1993, PASP, 105, 1342

  \bibitem[\protect\citeauthoryear{Stellingwerf}{1978}]{stelling} Stellingwerf B. F., 1978, ApJ, 224, 953

  \bibitem[\protect\citeauthoryear{Udalski, Kubiak \& Szymanski}{Udalski et al.}{1997}]{udalski1997} Udalski A., Kubiak M., Szyma\'{n}ski M., 1997, AcA, 47, 319

  \bibitem[\protect\citeauthoryear{Udalski et al.}{1999}]{udalski1999} Udalski A., Szyma\'{n}ski M., Kubiak M., Pietrzy\'{n}ski G., Soszy\'{n}ski I., Wo\'{z}niak P., \.{Z}ebru\'{n} K., 1999, AcA, 49, 201

  \bibitem[\protect\citeauthoryear{Westerlund}{1997}]{westerlund} Westerlund B. E., 1997, The Magellanic Clouds. Cambridge Univ. Press, Cambridge

  \bibitem[\protect\citeauthoryear{Whitelock, Marang \& Feast}{Whitelock et al.}{2000}]{whitelock2000} Whitelock P. A., Marang F., Feast M. W., 2000, MNRAS, 319, 728

  \bibitem[\protect\citeauthoryear{Whitelock et al.}{2003}]{whitelock2003} Whitelock P. A., Feast M. W., van Loon J. Th., Zijlstra A. A., 2003, MNRAS, 342, 86

  \bibitem[\protect\citeauthoryear{Wood \& Sebo}{1996}]{wood1996} Wood P. R., Sebo K. M., 1996, MNRAS, 282, 958

  \bibitem[\protect\citeauthoryear{Wood et al.}{1999}]{wood1999} Wood P. R. et al., 1999, IAU symp., 191, 151

  \bibitem[\protect\citeauthoryear{Wood}{2000}]{wood2000} Wood P. R., 2000, PASA, 17, 18

  \bibitem[\protect\citeauthoryear{Wood, Olivier \& Kawaler}{Wood et al.}{2004}]{wood2004} Wood P. R., Olivier E. A., Kawaler S. D., 2004, ApJ, 604, 800

  \bibitem[\protect\citeauthoryear{Wray, Eyer \& Paczy\'{n}ski}{Wray et al.}{2004}]{wray2004} Wray J. J., Eyer L., Paczy\'{n}ski B., 2004, MNRAS, 349, 1059

  \bibitem[\protect\citeauthoryear{\.{Z}ebru\'{n} et al.}{2001}]{zebrun2001} \.{Z}ebru\'{n} K., et al., 2001, AcA, 51, 317

\end{thebibliography}
\end{document}